\documentclass[amssymb,aps,showkeys,]{revtex4-2}

\usepackage{graphicx}
\usepackage{url}
\usepackage{multirow}
\usepackage{color}
\usepackage{amsmath}
\usepackage{subcaption}
\usepackage{pgfplots}

\usepackage{tikz}
\usetikzlibrary{shapes.geometric,arrows}
\usepackage{qcircuit}
\usepackage{braket}
\usepackage{adjustbox} 
\usepackage{booktabs}

\usepackage{geometry}
\usepackage{array}
\usepackage{multirow}

\usepackage[linesnumbered,ruled,vlined]{algorithm2e}

\usepackage{comment}

\includecomment{sectiontocompile}

\usepackage{pgfplots}
\pgfplotsset{compat=newest}
\usepgfplotslibrary{groupplots}
\usepgfplotslibrary{dateplot}

\renewcommand{\thesection}{\arabic{section}}
\renewcommand{\thetable}{\arabic{table}}
\renewcommand{\thesubsection}{\thesection.\arabic{subsection}}

\begin{document}

\title{A novel feature selection method based on quantum support vector machine}

\author{Haiyan Wang}
\email{haiyan.wang@asu.edu}
\affiliation{School of Mathematical and Natural Science,  Arizona State University, Phoenix, AZ 85069, USA}


\begin{abstract}

Feature selection is critical in machine learning to reduce dimensionality and improve model accuracy and efficiency. The exponential growth in feature space dimensionality for modern datasets directly results in ambiguous samples and redundant features, which can severely degrade classification accuracy. Quantum machine learning offers potential advantages for addressing this challenge. In this paper, we propose a novel method, quantum support vector machine feature selection (QSVMF), integrating quantum support vector machines with multi-objective genetic algorithm. QSVMF optimizes multiple simultaneous objectives: maximizing classification accuracy, minimizing selected features and quantum circuit costs, and reducing feature covariance. We apply QSVMF for feature selection on a breast cancer dataset, comparing the performance of QSVMF against classical approaches with the selected features. Experimental results show that QSVMF achieves superior performance. Furthermore, The Pareto front solutions of QSVMF enable analysis of accuracy versus feature set size trade-offs, identifying extremely sparse yet accurate feature subsets. We contextualize the biological relevance of the selected features in terms of known breast cancer biomarkers. This work highlights the potential of quantum-based feature selection to enhance machine learning efficiency and performance on complex real-world data.

\end{abstract}


\maketitle


\section{Introduction}\label{sec1}
Feature selection is a common way to minimize the problem of excessive and irrelevant features, which is one of the most fundamental problems in machine learning and pattern recognition. Overcoming the curse of dimensionality is one of the biggest challenges in building an accurate predictive machine learning model from high dimensional data. For a feature selection problem with m features, the objective of feature selection should be to achieve the highest classification accuracy by finding an optimal feature subset. The number of possible feature subsets is an optimal solution is NP-hard \cite{Pudjihartono_2022,Gopal_2021,Altarabichi_2023,Saeys_2007}.

Quantum machine learning has emerged as a highly promising application of quantum technologies. In the search for quantum advantage, the quantum support vector machines, stand out as a powerful approach, capturing the interest of diverse research groups \cite{BYH22,GGC22,LAT21,MBS22,PQW20,RML14,SCH21,SYG20, BWP17,SSP14,Schuld2022, huang2021,Di2023,HCT19,MS21}. In the work presented by \cite{HCT19}, innovative strategies were introduced to construct quantum kernels for support vector machines (SVM). This method utilizes classical data and transforms it into the quantum state space through a quantum feature map, enabling the conversion of the dataset from its original low-dimensional real space to a higher-dimensional quantum state space, commonly known as the Hilbert space.  More recently, quantum computing has emerged as a promising platform fort tackling computationally expensive combinatorial optimization tasks such as feature selection\cite{Mucke_2023,wanglu_2023,Turati_2022, Zoufal_2022}.

Grounded in the principles of natural selection, genetic algorithms emerge as a valuable tool for feature selection \cite{Sowan_2022, Altarabichi_2023}. Recently, genetic algorithms have found application in support vector machines \cite{Ji2020}. Genetic algorithms have been extensively applied in quantum computing \cite{Lahoz-Beltra2016, Acampora2021, CYH16} and quantum support vector machines \cite{Kavitha2022, ARG21, CC2022}. The concept of using genetic algorithms for the automatic generation of feature maps in quantum support vector machines was first introduced by \cite{ARG21}. In addition, \cite{baran2021, CSU20} proposed algorithms for automatically generating quantum circuits using multi-objective genetic algorithms. These algorithms offer a means to optimize gate or circuit structures, addressing common challenges such as local minima and barren plateaus \cite{li2017, lamata2018, chivilikhin2020}. 

In this paper, we introduce a new feature selection method, quantum support vector machine feature selection (QSVMF), based on quantum support vector machine (QSVM) and a multi-objective genetic algorithm (specifically the Non-dominated Sorting Genetic Algorithm (NSGA-II)). We use five fitness functions that simultaneously maximizes classification accuracy while minimizing, the number of selected features, the gate costs of the quantum feature map's circuit and the covariance of selected features. The inclusion of the minimization of quantum gate costs and the feature covariances helps reducing high computational cost which is a major drawback of using genetic algorithms. In particular, we separate the fitness functions for local gates and non-local (controlled-NOT (CNOT)) gates for balanced quantum gates for entanglement, which complements the previous works \cite{ARG21,CC2022} where their formulation of fitness functions has primarily resulted in the suppression of entanglement gates in the design. We apply QSVMF to the feature selection of the breast cancer dataset and compare QSVMF with SelectKBest from scikit-learn library and classical methods. QSVMF outporforms in the most cases.  
 
In addition, the use of NSGA-II in this paper allows us to examine the Pareto front solutions for an interesting analysis of the trade-off between number of features and accuracy using the QSVMF method. In particular, we include the minimization of the total covariance of selected features to reduce the redundance of features. As a result, we identify the feature subsets with the minimum number of features for each qubit setting. Despite having very few features, these compact QSVMF feature subsets can still achieve good accuracy. This demonstrates another key advantage of the QSVMF approach - its ability to pare down the feature space to a very sparse set of highly informative features critical for prediction. The compact high-accuracy feature subsets could help enable more interpretable and efficient predictive modeling.

Finally, we conceptualize that the features chosen by QSVMF align with established medically relevant imaging biomarkers for differentiating between benign and malignant breast lesions. A future comprehensive study on diverse datasets will fully justify the advantages of QSVMF across various data distributions and complexities.

Several studies have explored quantum feature selection approaches, including those based on a quadratic unconstrained binary optimization (QUBO) problem \cite{Mucke_2023}, Hamiltonian encoding and a ground state \cite{wanglu_2023}, quantum approximate optimization algorithm (QAOA) \cite{Turati_2022} and variational quantum optimization with black box binary optimization \cite{Zoufal_2022}. To the author's knowledge, this paper marks the first application of QSVM and NSGA-II for quantum feature selection.

\section{Feature Selection}
\label{sec:trad_fs}
Feature selection is a technique aimed at enhancing training efficiency. It chooses a subset of features for training a model, with the primary objectives of expediting model training, minimizing memory requirements, and mitigating overfitting risks. Additionally, since not all features may contribute significantly to the training process, feature selection plays a pivotal role in enhancing model interpretability and promoting generalization.

The primary classical feature selection categories can be broadly classified into three groups: 1) filters; 2) wrappers; and 3) embedded methods, as discussed in \cite{Pudjihartono_2022}. Filter methods employ statistical techniques to independently evaluate the relationships between features and the output, eliminating features that fail to meet predefined thresholds. Wrapper methods evaluate numerous feature subset models, selecting the most effective features that maximize overall model accuracy. Embedded methods seamlessly integrate feature selection into the classifier algorithm. During training, the classifier adjusts its parameters to determine the importance of each feature for optimal classification accuracy. Embedded methods strive to strike a balance between the characteristics of filter and wrapper methods.

In this paper, we propose a new embedded method, QSVMF, for feature selection based on high-dimensional NSGA-II (Non-dominated Sorting Genetic Algorithm II) and quantum support vector machine. The new method utilizes a SVM with quantum kernels to maximize prediction accuracy and NSGA-II to reduce computational cost by minimizing complexity of quantum circuits and correlation of selected features.  As a result, QSVMF outperforms many classical methods for feature selection of the breast cancer dataset. 

\section{Quantum Support Vector Machine}\label{sec2}

\subsection{Support vector machine}\label{sec2-1}

Support vector machines (SVMs) represent robust supervised learning models employed for classification tasks. The objective is to identify the optimal hyperplane that effectively maximizes the separation between two classes. In mathematical terms, considering a labeled training set ${(x_i, y_i)}$ where $x_i$ denotes the features and $y_i$ signifies the binary labels, this objective is mathematically formulated as:

\begin{align*}  
&\text{minimize} \quad ||w||^2 \\
&\text{subject to} \quad y_i(w \cdot x_i + b) \geq 1, \quad \forall i
\end{align*}
Here, $w$ represents the normal vector to the hyperplane, and $b$ is the bias term. The constraints are in place to guarantee the maximization of the margin. The points $x_i$ closest to the hyperplane are referred to as support vectors. This primal optimization can be rephrased in the Lagrangian dual form:

\begin{equation} 
\begin{aligned}
\max_\alpha & L(\alpha) = \sum_{i=1}^{n}\alpha_i - \frac{1}{2}\sum_{i=1}^{n}\sum_{j=1}^{n} \alpha_i\alpha_jy_iy_jx_i \cdot x_j\\
\text{s.t.} & \sum_{i=1}^{n} \alpha_i y_i=0, \quad \alpha_i \geq 0
\end{aligned}
\end{equation}

Given a function $\Phi$ that maps $x$ to a higher-dimensional space, this dual form allows for the utilization of kernels, denoted as $K(x_i, x_j) = \Phi(x_i)\cdot\Phi(x_j)$, enabling the implicit evaluation of inner products. This technique, commonly known as the "kernel trick," plays a crucial role in extending the capabilities of SVMs beyond linear problems. In essence, kernels facilitate the transformation of data into higher-dimensional feature spaces, where non-linear problems become linearly separable.

\subsection{Quantum kernel method}\label{sec2-2}
The Quantum Kernel Estimator described in \cite{HCT19} provides effective strategies for constructing a quantum kernel for a support vector machine. A major challenge in the construction of quantum kernel is encoding classical data into quantum states for quantum computing. \cite{HCT19} achieves this encoding by mapping data into the quantum state space through a quantum feature map. The choice of feature map is critical, often dependent on the dataset's characteristics to be classified. The feature map's importance goes beyond utilizing the quantum state space as a feature space. It also involves how the data gets mapped into this high-dimensional space.

By employing quantum circuits, a data point $\mathbf{x}=(x_j) \in \mathbb{R}^n$ undergoes transformation into an n-qubit quantum feature state, denoted as $|\Phi(\mathbf{x}\rangle\langle\Phi(\mathbf{x})|$. This naturally gives rise to the definition of the kernel as $K(\mathbf{x},\mathbf{z})= | \langle \Phi (\mathbf{x}) \vert \Phi (\mathbf{z}) \rangle |^2 $. The incorporation of the kernel method into quantum computing seamlessly occurs through the utilization of a unitary operator $\mathcal{U}_{\Phi(\mathbf{x})}$ applied to the initial state $\vert 0 \rangle ^n$ as $\vert \Phi (\mathbf{x}) \rangle = \mathcal{U}_{\Phi(\mathbf{x})} \vert 0 \rangle ^n$. This approach maps data points into the quantum Hilbert space ~\cite{SK19}, thereby capitalizing on the significant advantages offered by quantum computing.

\begin{equation}\label{K_circuit}
    K(\mathbf{x},\mathbf{z}) = |\langle\Phi(\mathbf{x})|\Phi(z)\rangle|^2
    = |\langle 0^n|\mathcal{U}_{\Phi(\mathbf{x} }^\dagger)\mathcal{U}_{\Phi(\mathbf{z})}|0^n\rangle|^2
\end{equation}\medskip

In the work by \cite{HCT19}, several well-established feature maps, such as ZFeatureMap and ZZFeatureMap, are introduced. A customization of quantum feature maps involves the utilization of more Pauli gates. As an illustration, a representative form of our customized quantum feature map could assume the following structure:

$$
\mathcal{U}_{\Phi(\mathbf{x})}=\left(\exp \left(i \sum_{j, k} \phi_{\{j, k\}}(\mathbf{x}) Z_{j} \otimes Z_{k}\right) \exp \left(i \sum_{j} \phi_{\{j\}}(\mathbf{x}) P_{j}\right) H^{\otimes n}\right)^{d}
$$ 
Here, $P_j$ represents one of the rotation gates $I, X, Y, Z$. The quantum circuit for $d=1$ is depicted in Table \ref{he_moon23} for $n=2$ qubits. The coefficient $\phi_{S}(x)$ is computed, with $S$ denoting a set comprising independent qubits or qubit pairs, as illustrated below
\begin{equation}
\phi_S: x \mapsto
\begin{cases}
x_{i} & \text{if } S = \{i\} \\
2x_{i}x_{j} & \text{if } S = \{i, j\}
\end{cases}
\label{eq56}
\end{equation}
to encode the data.  

\begin{table}[h]
\[
\begin{array}{c}
\Qcircuit @C=1em @R=.7em {
\lstick{\ket{0}} & \gate{H} & \gate{R} & \ctrl{1} &  \qw  & \ctrl{1}&\qw \\
\lstick{\ket{0}} & \gate{H} & \gate{R} & \targ & \gate{R} & \targ & \qw
}
\end{array}
\]
\caption{Quantum circuit for $n=2$ qubits }  
\label{he_moon23} 
\end{table}
\noindent The circuit illustrated in Table \ref{he_moon23} consists of a layer of Hadamard gates $H^{\otimes 2}$, followed by an additional layer of single-qubit gates $R=e^{i \phi_{{j}}(\mathbf{x}) P_{j}}$. These $R$ gates are defined by a set of angles $\phi_{{j}}(\mathbf{x})$, each corresponding to a unique axis, and $\phi_{{j}}(\mathbf{x})$ is dependent on the feature data. The diagonal entangling gate $e^{i \phi_{{0,1}}(\mathbf{x}) Z_{0} \otimes Z_{1}}$ is characterized by an angle $\phi_{{0,1}}(\mathbf{x})$ and can be realized with two CNOT gates and one $R=e^{i \phi_{{0,1}}(\mathbf{x}) Z_{1}}$ gate, as illustrated in Table \ref{he_moon23} \cite{HCT19}. In this context, the CNOT gate establishes entanglement among qubits, specifically acting on two qubits, and is represented by the matrix:
$$
\mathrm{CNOT}=\left[\begin{array}{llll}
1 & 0 & 0 & 0 \\
0 & 1 & 0 & 0 \\
0 & 0 & 0 & 1 \\
0 & 0 & 1 & 0
\end{array}\right]
$$
The first qubit functions as the control qubit, while the second qubit is assigned as the target qubit. If the control qubit is in the state $|0\rangle$, the target qubit maintains its original state. On other hand, when the control qubit is in the state $|1\rangle$, the target qubit undertakes a bitwise flip. The classical equivalent of the CNOT gate aligns with the reversible XOR gate.

Creating a suitable kernel function for diverse datasets poses a significant challenge. In our investigation, we employ a genetic algorithm to execute a global optimization strategy for refining the structure of the quantum circuit. This approach has proven to be effective in overcoming common challenges, such as barren plateaus, encountered in traditional optimization methods. Consequently, it produces high-quality learning kernels \cite{ARG21, GA11}.



\section{Multi-Objective Genetic Algorithm} \label{sec3}

Genetic algorithms are optimization methodologies inspired by the principles of evolution. These algorithms navigate a solution space by iteratively refining a population of individuals. Across successive generations, genetic operations guide the selection of offspring with the aim of improving one or more objectives.  This evolutionary process concludes with the identification of the individuals to optimize fitness functions within a vast configuration space.

When generating new solutions, the process involves selecting "parent" solutions from a predefined pool. Utilizing mechanisms such as crossover and mutation, a "child" solution is created, inheriting traits from its "parents." New parent pairs are chosen for each child, and this process is repeated until a new population of the desired size is formed. The effectiveness and applicability of a genetic algorithm rely significantly on the choice of genetic operations. Among these operations, the selection operator is crucial as it determines a subset of the current population to create a new generation through crossover and mutation. Mutation involves randomly altering information in selected individuals, allowing for exploration of distant regions in the solution space. On the other hand, crossover facilitates more radical exploration by exchanging genetic information between two individuals.

To guarantee efficiency and convergence, genetic algorithms incorporate early stopping conditions to evaluate whether the evolutionary process has achieved its goals. These conditions may include closely examining fitness convergence or saturation, maintaining a minimum accuracy threshold for continuation, or defining a maximum generation count. By using these essential components, genetic algorithms demonstrate their value in tackling complex optimization problems across various domains

For many problems, an evolutionary multi-objective optimization (EMO) may be more effective, which involves multiple objective functions that require either minimization or maximization  \cite{ADB14, Miettinen1999,Deb2002}. Multi-objective optimization, like single-objective problems, may also have various constraints that all feasible solutions.  



\tikzstyle{startstop} = [rectangle, rounded corners, 
minimum width=3cm, 
minimum height=1cm,
text centered, 
draw=black, 
fill=red!30]

\tikzstyle{description} = [rectangle, 
minimum width=2cm, 
minimum height=1cm, 
text centered, 
text width=2cm, 
draw=black, 
fill=orange!30]

\tikzstyle{decision} = [diamond, 
minimum width=2cm, 
minimum height=1cm, 
text centered, 
text width=2cm, 
draw=black, 
fill=green!30]
\tikzstyle{arrow} = [thick,->,>=stealth]

\begin{figure}
\centering

\tikzstyle{process} = [rectangle, 
minimum width=3cm, 
minimum height=1cm, 
text centered, 
text width=3cm, 
draw=black, 
fill=blue!20]

\tikzstyle{processY} = [rectangle, 
minimum width=3cm, 
minimum height=1cm, 
text centered, 
text width=3cm, 
draw=black, 
fill=yellow!10]

\begin{tikzpicture}[node distance=0.5cm]
\node (1) [process] {Initialize population};
\node (2) [process, right of=1, xshift=5cm]  {Perform Crossover and Mutation};
\node (4) [process, below of=2, yshift=-1cm] {Combine Parent and Offspring Population};
\node (5) [process, below of=4,yshift=-1cm]  {Evaluate Fitness Functions};
\node (6) [process, below of=5, yshift=-1cm] {Non-dominated Sorting and Rejection};
\node (7) [process, below of=6, yshift=-1cm] {Crowding Distance Sorting};
\node (8) [process, below of=7, yshift=-1cm] {Check Stopping Criteria};
\node (9) [process, left of=8, xshift= -5cm] {Identif Pareto Fronts};
\node (10) [process, right of=8, xshift=5cm] {Select Sruvivors for Next Generation};
\node (11) [processY, left of=5, xshift=-5cm] {Apply QSVM};


\draw [arrow] (1)--(2) ;
\draw [arrow] (2)--(4) ;
\draw [->](4) -- (5);
\draw [arrow] (5) -- (6);
\draw [arrow] (6) -- (7);
\draw [arrow] (7) -- (8);
\draw [arrow] (8) -- node[anchor=east] {yes} (9);
\draw [arrow] (8) -- node[anchor=south] {no} (10);
\draw [arrow] (10) |- (2);

\draw [arrow,dashed] (5) -- (11);
\draw [arrow,dashed] (11) -- (5);

\end{tikzpicture}
  \caption{Non-dominated Sorting Genetic Algorithm (NSGA-II)}
  \label{NSGA-II}
\end{figure}
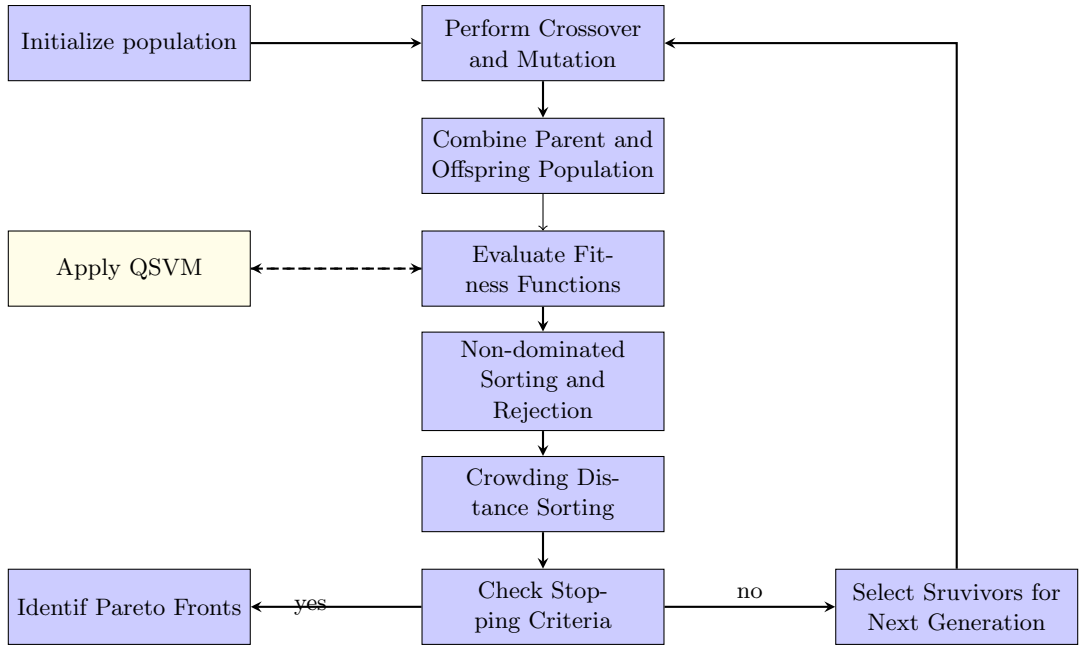

Optimal solutions in multi-objective optimization are described by the mathematical concept of partial ordering, expressed through the term "domination."  This paper only addresses unconstrained optimization problems, which do not involve any equality, inequality, or bound constraints. We say that solution $\mathbf{x}^{(1)}$ dominates another solution, $\mathbf{x}^{(2)}$, only if both of the following conditions hold true:

\begin{enumerate}
  \item The solution $\mathbf{x}^{(1)}$ is no worse than $\mathbf{x}^{(2)}$ in all objectives (it holds for each component of the vector.).  
  \item The solution $\mathbf{x}^{(1)}$ is strictly better than $\mathbf{x}^{(2)}$ in at least one objective ( it holds for at least one component of the vector)
\end{enumerate}

Points that persist without being dominated by any other members within a set are labeled as non-dominated points. A notable characteristic of such points is their trade-off relationship, where improvements in one objective affect trade-offs in at least one other objective. This trade-off property observed among non-dominated points is interesting as it promotes the exploration of a diverse array of such points before arriving at a final decision. The Pareto front represents the collection of such non-dominated solutions, forming a frontier where any improvement in one objective necessitates a trade-off in another. 

The Non-dominated Sorting Genetic Algorithm (NSGA-II) emerged as a response to the deficiencies observed in early evolutionary algorithms, particularly the absence of elitism, as discussed in \cite{Deb2002}. It has become a widely used technique in the field of Evolutionary multi-objective optimization.

As we structure our fitness functions to encompass the minimization of gate costs and the total covariances of selected features, the Pareto fronts derived for QSVMF provide insights into the strategy of balancing high accuracy with minimal gate costs and selected features. This facilitates the analysis of cost-effective trade-off strategies. A more thorough investigation into the multiple-dimensional Pareto fronts and their projections onto low-dimensional spaces could offer a deeper understanding of the significant impact of quantum feature maps on feature selection.

\section{Quantum Kernel for Support Vector Machine}\label{sec4}

\subsection{Genetic quantum feature map}\label{featureMap}
Utilizing the approach outlined in \cite{ARG21} to generate quantum feature maps through a multi-objective optimization process, this paper employs the NSGA-II algorithm to automatically generate quantum feature maps for support vector machines. As previously mentioned, the NSGA-II algorithm incorporates elitism and other features to ensure diversity within the Pareto set. It searches for circuits that, when trained using a support vector machine, maximize accuracy while simultaneously minimizing circuit complexity and feature redundancy. 

Gate costs are quantified in terms of local gates, including Hadamard gates and rotation gates, as well as CNOT gates.  In this study, we introduce a refined approach by separate CNOT gates from local gates and incorporating additional optimization functions into the circuit generation process. Building upon the encoding method established in \cite{ARG21,CC2022}, our modifications aim to create a quantum circuit with minimal gate costs and covariances of features while maximizing accuracy. Assume that the number of qubits is $N$. The encoding scheme starts with the number of bits equal to the number of features to indicating if a feature is selected, and then $N$ Hadamard gates, followed by $N$ bits indicating whether if a rotation is applied to each qubit, and $\binom{N}{2}$ bits for the number of potential CNOT gates and an additional two bits to denote rotations with respect to $X$, $Y$, or $Z$. The last two bits are for repetition information. The size of the encoding bit string is given by
$$
\text{number of features} + N  + \binom{N}{2} + 4  
$$ 
Such scheme includes possible Hadamard entanglement, achievable through two rotation gates and a CNOT gate, as illustrated in Table \ref{he_moon23}. We incorporate the input data into the encoding. If the feature number is the same as the number of qubits, ensuring that the $i$-th gene operating on the qubits of the quantum register depends on the $i$-th variable from the input data $\mathbf{x} \in \mathbb{R}^n$, as expressed in Equation (\ref{eq56}). If the feature number is larger than the number of qubits, we randomly select the number of feature as the number of qubits for embedding the feature data to the feature maps. If the feature number is less than the number of qubits, we randomly copy some features to match the number of qubits for embedding the feature data to the feature maps.

The genetic algorithm initiates the population randomly, and each individual will associate and quantum circuit and undergoes fitness evaluation. Utilizing the training dataset, the circuit assesses kernel matrices, which are integrated into SVM algorithms to compute the fitness function. Individuals with superior fitness are more likely to undergo genetic operations, including mutation, crossover, and selection, leading to the creation of a new generation of individuals or quantum circuits. This iterative process continues until convergence criteria, including comparison of accuracy and the number of generation, are met. To illustrate the our encoding scheme, we consider a dataset with four features, three qubits with 14 bits per individual in the population, as depicted in Figure \ref{encode-1}.  Figure \ref{encode} is the corresponding circuits of the quantum feature map.

\tikzstyle{s} = [rectangle, rounded corners, 
minimum width=3cm, 
minimum height=0.5cm,
text centered, 
draw=black, 
fill=red!30]

\tikzstyle{description} = [rectangle, 
minimum width=2cm, 
minimum height=1cm, 
text centered, 
text width=2cm, 
draw=black, 
fill=orange!30]

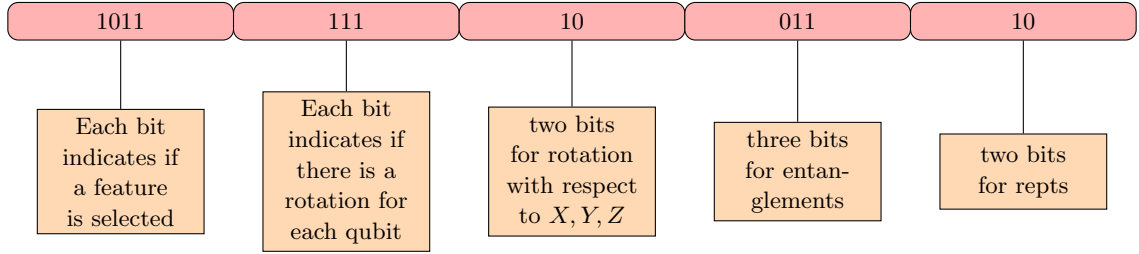
\begin{figure}[htbp]
\centering
\begin{tikzpicture}[node distance=2cm]
\node(1)[s,right of=1, xshift=2cm]{1011};
\node(2)[s,right of=1, xshift=1cm]{111};
\node(3)[s,right of=2, xshift=1cm]{10};
\node(4)[s,right of=3, xshift=1cm]{011};
\node(5)[s,right of=4, xshift=1cm]{10};

\node(1e)[description,below of=1]{Each bit indicates if a feature is selected};
\node(2e)[description,below of=2]{Each bit indicates if there is a rotation for each qubit};
\node(3e)[description,below of=3]{two bits for rotation with respect to $X, Y, Z$};
\node(4e)[description,below of=4]{three bits for entanglements};
\node(5e)[description,below of=5]{two bits for repts};

\draw [-] (1) -- (1e);
\draw [-] (2) -- (2e);
\draw [-] (3) -- (3e);
\draw [-] (4) -- (4e);
\draw [-] (5) -- (5e);
\end{tikzpicture}
\caption{Example for the encoding scheme for 4 features and 3 qubits: the binary bits are [1,0,1,1,1, 1, 1, 1, 0, 0, 1, 1, 1, 0], the first four bits indicate that features 0,2,3 are chosen,  the next three bits for 3 rotation gates, and next two bits indicates the rotation with respect to $X, Y, Z$, followed 3 bits for possible entanglements,  the last two bits are for repetitions.}
\label{encode-1}
\end{figure}

\begin{figure}[htbp]
    \centering
    \includegraphics[width=0.8\textwidth]{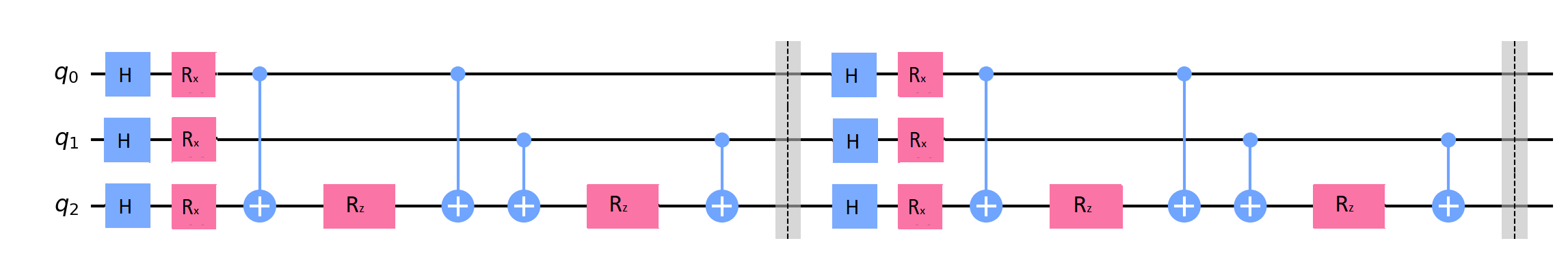}
    \caption{Quantum circuits for Figure \ref{encode-1}}
    \label{encode}
\end{figure}

\subsection{Fitness functions}\label{sec3-2}

fitness functions are the he objective or cost function for measuring the effectiveness of QSVMF, taking a solution as input and providing its fitness score. Introducing five fitness functions inspired by \cite{ARG21} below (\ref{FitnessFunc}), our goal is two-fold: to maximize accuracy and minimize gate costs and covariance of selected features.  Recognizing the distinct costs associated with different gates \cite{ARG21, CC2022, LLK06}, we differentiate CNOT gates from local gates, leading to the introduction of two fitness optimization functions for gate costs. Gate costs are computed by adding the number of gates used in the operations required for quantum computer implementation.  
We choose not to integrate weights into the objective functions, as the introduction of such weights might have a potential impact on the optimization process in a manner that is intentionally undesirable. In our pursuit of this optimization, we conduct a evaluation using K-fold cross-validation.  Here accuracy represents the mean accuracy achieved on the provided test data and labels over the k-fold cross-validation.

\begin{equation}\label{FitnessFunc} 
\left\{
\begin{array}{lr}
\text{(maximize)} \text{Fitness 1} = \text{prediction accuracy}\\ 
\text{(minimize)} \text{Fitness 2 } =\text{ local gates}=\text{Rgate} + \text{Hgate} \\
 \text{(minimize)} \text{Fitness 3 } = \text{CNOT gate}\\
 \text{(minimize)} \text{Fitness 4 } = \text{feature count}\\
 \text{(minimize)} \text{Fitness 5 } = \text{feature variance}
\end{array}
\right. 
\end{equation}


\section{Feature Selection with QSVMF}
\label{sec:VarQFS_application}

\subsection{Breast cancer dataset}

Our analysis uses the breast cancer dataset retrieved from Sklearn \cite{PVG11, Zwitter1998}. This dataset encompasses measurements of breast tissue obtained through medical imaging techniques, featuring various metrics. The primary goal is to determine the nature of a tumor, distinguishing between benign (non-cancerous) and malignant (cancerous and hazardous) tumors. The breast cancer dataset from scikit-learn, consists of 569 samples, incorporating 30 real, positive features. These features include attributes related to cancer masses, including mean radius, mean texture, mean perimeter, and more. Among these samples, 212 instances are labeled as "malignant," while 357 instances are labeled as "benign." We opt not to employ principal component analysis (PCA) for dimension reduction for intensifying the challenge for achieving high accuracy.

\subsection{QSVMF Procedure}

\begin{figure}[htbp]
\centering

\tikzstyle{processA} = [rectangle, 
minimum width=1cm, 
minimum height=4cm, 
text centered, 
text width=1.8cm, 
draw=black, 
fill=yellow!30]

\begin{tikzpicture}[node distance=1cm]

\node (2) [processA, right of=1, xshift=1.7cm] {Data preprocess (K-fold, population encoding,..)};
\node (3) [processA, right of=2,xshift=1.7cm] {NSGA-II process (Crossover, Mutation)}; 
\node (3a) [processA, right of=3,xshift=1.7cm] {Feature selection according to encoding scheme}; 
\node (4) [processA, right of=3a,xshift=1.7cm] {Compute quantum kernel};
\node (6) [processA, right of=4, xshift=1.7cm] {Compute accuracy using QSVM};
\node (8) [processA, right of=6, xshift=1.7cm] {NSGA-II process (sorting and refining)};
\node (9) [decision, below of=4, xshift= -4.5cm,yshift= -3cm] {K-fold criteria};

\node (11) [decision, above of=4, xshift= 1.5cm, yshift= 3cm] {NSGA-II repetition};
\node (12) [processA, below of=9, xshift= 3cm, yshift= -3cm] {Extract selected features from Pareto fronts};
\node (13) [processA, right of=12, xshift= 2.7cm] {Perform predictions for comparison};

\draw [->] (2) -- (3);
\draw [->] (3) -- (3a);
\draw [->] (3a) -- (4);
\draw [arrow] (4) -- (6);
\draw [arrow] (6) -- (8);
\draw [arrow] (8) |- (9);
\draw [arrow] (12) -- (13);

\draw [arrow] (8) |- (11);
\draw [arrow] (11) -| (3);

\draw [arrow] (9) -| node[anchor=north] {no} (2);
\draw [arrow] (9) |- node[anchor=south] {yes} (12);

%

\end{tikzpicture}
  \caption{Experimental procedure with NSGA-II and quantum feature map}
  \label{proc11}
\end{figure}
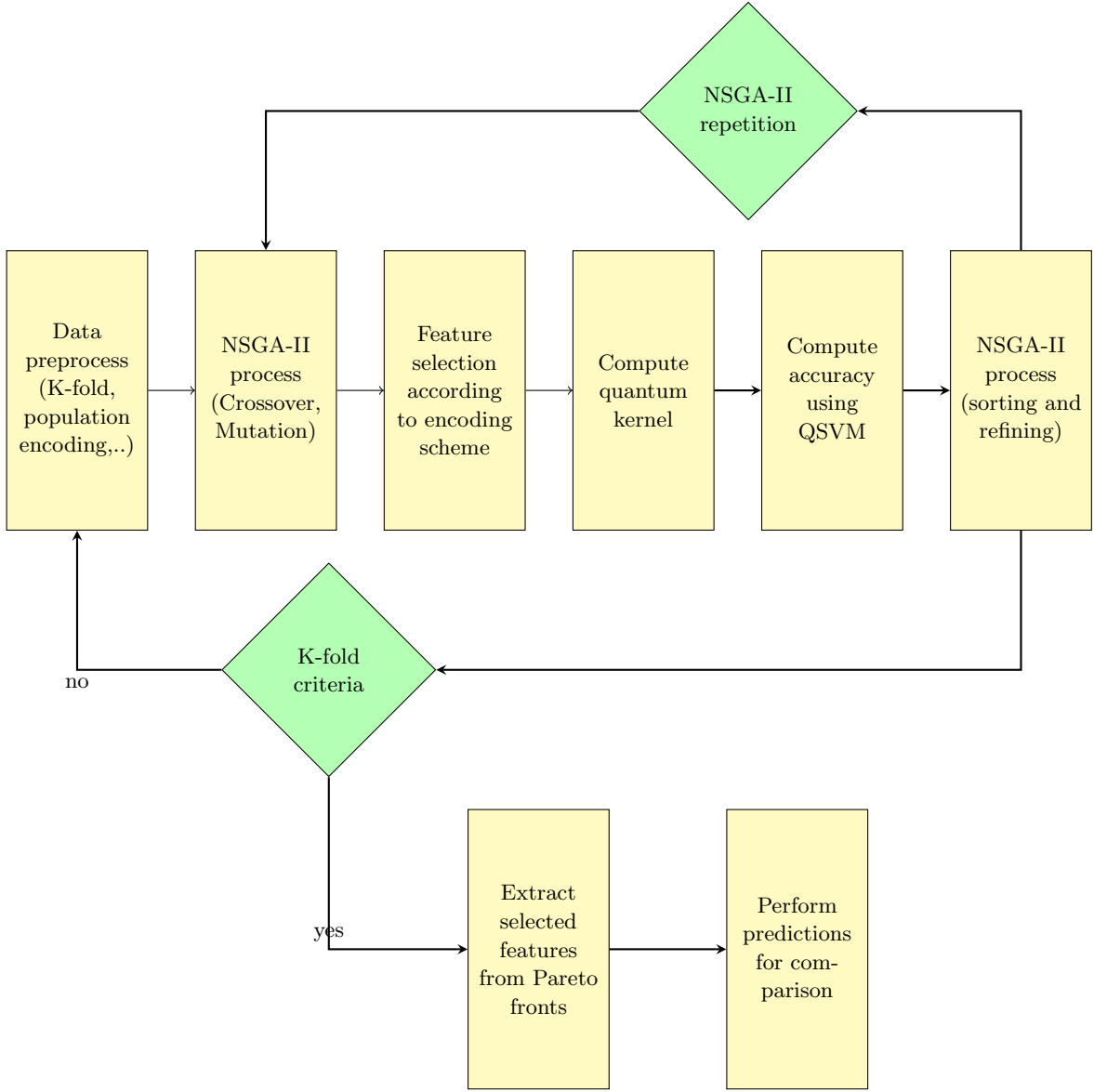

The experiment utilizes high-dimensional NSGA-II (Non-dominated Sorting Genetic Algorithm II) and quantum feature mapping to enhance accuracy. After encoding each individual in the population for quantum feature maps, genetic algorithm operators such as selection, crossover, and mutation are applied to generate the subsequent generation of individuals. Subsequently, we calculate the kernel of a quantum SVM while training a classifier with the provided training dataset. In each iteration, we assess the classifier's accuracy using the test dataset and evaluate the effective size of the circuit in terms of the number of local and entangling gates it incorporates. Special precautions are taken to prevent overfitting by implementing K-fold cross-validation and comparing the performance with the baseline model. The procedure is structured into a sequence of interconnected steps depicted by distinct rectangular blocks, as illustrated in Figure \ref{proc11}, and can be further described as follows.

\begin{itemize}
    \item \textbf{Data Preprocess}: The first step comprises data preprocessing utilizing a K-fold strategy, which divides the dataset into subsets for both training and testing. Additionally, individuals within populations are encoded for quantum feature maps

    \item \textbf{NSGA-II Process (Crossover, Mutation)}: It encompasses various genetic operators such as crossover and mutation, designed to evolve a population of potential solutions to improve their quality with respect to multiple objectives.
    
    \item \textbf{Feature selection According to Encoding Scheme}: Features are selected Before computing kernel matrix according to the encoding scheme. 
        
    \item \textbf{Construct Quantum Feature Map and Compute Kernel Matrix}: 
    We construct a quantum feature map, a pivotal component of SVM that transforms classical data into a quantum state. The process also includes the computation of a kernel matrix to capture data relationships.
    
    \item \textbf{Classical SVM (quantum kernel) to Compute Accuracy}: A classical SVM is utilized with a quantum kernel to calculate the accuracy of the model's predictions, enhancing the efficacy of this potent classical method in machine learning. 

    \item \textbf{NSGA-II Process (Sorting and refining)}: The individuals are sorted based on their dominance, potentially deleting and refining the solutions generated earlier.

    \item \textbf{Extract Pareto Fronts and the most frequent features}: The final step aims to obtain $K$ Pareto fronts and the most  $m$ frequnt features from the resulting Pareto fronts associated with the highest accuracy, where $m$ is the average of the numbers of features of each K-fold cross validation. 

     \item \textbf{Perform predictions with both classical and quantum classifiers for comparison}: With the selected feature list, we perform various predictions with both classical and quantum classifiers for comparison.
\end{itemize}

With our training and testing datasets prepared, we follow the previously outlined procedure to compute the training and testing kernel matrices. We initiate the process with an initial random population of 100 individuals, represented as binary string. Each string comprises $\text{number of features}+N+\binom{N}{2} +4$ bits, where $N$ denotes the number of qubits. Employing a 20\% crossover and mutation rate, we aim for a more realistic accuracy by utilizing the 5-fold cross-validation method, resulting in five distinct training and test subsets. Subsequently, we average accuracy and other outputs for a comprehensive assessment. We employ a combination of generation number and accuracy as a stopping criterion. These quantum kernels can be seamlessly integrated into classical kernel methods. Notably, NSGA-II typically runs around 100 generations to achieve higher accuracy compared to classical methods.

\begin{algorithm}
\caption{QSVMF pseudo code}

\SetAlgoLined
\KwData{training data X, labels y, number of qubits}
\KwResult{selected features, comparsions}
\While{k-lold validation not done}{
\While{NSGA-II repetition not done}{
initialize population;
\For{each generation}{
Crossover and Mutation;\\
Construct Quantum Feature Map and Compute Kernel Matrix;\\
Use classical SVM with quantum kernel to compute accuracy and others;\\
update fitness;\\
Sorting individuals;\\
Rejection and Crowding Distance Sorting;
}
}
}
Obtain Pareto front solutions; compute accuracy with 5-fold cross validation;
get the most $m$ frequent selected features from the $k$ Pareto fronts associated with the highest accuracy, where $m$ is the average of the numbers of features;\\  
retrain final QSVM for comparison with other classical classifiers. \\
\Return selected features; comparisons 
\end{algorithm}

This multi-stage procedure is designed to achieve optimal solutions for complex optimization tasks while leveraging quantum computing principles and features an integrated approach that combines classical and quantum techniques. It is a sophisticated and comprehensive methodology for addressing challenging optimization and feature selection problems.

\subsection{Comparisons of QSVMF and SelectKBest }

SelectKBest is a feature selection method provided by the scikit-learn library in Python. The SelectKBest function takes two main arguments: score-func, which specifies the scoring function to use, and k, which specifies the number of top features to select. It selects the k highest scoring features based on a scoring function.  In this paper, the chi2 and f-regression scoring functions from scikit-learn were used for SelectKBest. The number $k$ of features selected by SelectKBest was set to be the same as the number of features selected by the QSVMF method. This allows a direct comparison between the classical SelectKBest and the quantum-based QSVMF in terms of their selected feature subsets. SelectKBest provides a simple baseline for feature selection that can be useful for benchmarking more advanced techniques like QSVMF.

\begin{table}[htbp]
    \centering
    \begin{adjustbox}{max width=15cm} 
        \begin{tabular}{p{3cm}lccccccccc}
            \toprule
            \multirow{2}{*}{ \parbox{3cm}{\raggedright{Feature Selection \\Method}}} & \multicolumn{9}{c}{qubits} \\
              \\[-1.2em]
                      \midrule 
            & 2   & 3   & 4   & 5   & 6   & 7   & 8   & 9   & 10  \\
            \midrule
            QSVMF   & [13,23] & [20,22] & [3,13,21] & [13,17,22,29] & [11,20,21,22] & [1,13,18,22,29] & [18,21,22,24] & [1,11,14,20,24] & [0,1,11,12,20,24,25]   \\
            \midrule
           \parbox{3cm}{\raggedright{SelectKBest \\(chi2)}}& [3, 23] & [3, 23] & [3, 13, 23] & [3, 13, 22, 23] & [3, 13, 22, 23] & [2, 3, 13, 22, 23] & [3, 13, 22, 23] & [2, 3, 13, 22, 23] & [0, 2, 3, 13, 20, 22, 23] \\
           \midrule
           \parbox{3cm}{\raggedright{SelectKBest \\(f-regression)}}& [22, 27] & [22, 27] & [7, 22, 27] & [7, 20, 22, 27] & [7, 20, 22, 27] & [2, 7, 20, 22, 27] & [7, 20, 22, 27] & [2, 7, 20, 22, 27] & [0, 2, 7, 20, 22, 23, 27]\\
            \bottomrule
        \end{tabular}
    \end{adjustbox} 
    \caption{Features selected by QSVMF and SelectKBest} 
    \label{table_featurelist}
\end{table}

Table \ref{table_featurelist} compares the feature subsets selected by QSVMF and the classical SelectKBest method with chi2 and f-regression scoring functions. The number of qubits used for QSVMF ranges from 2 to 10, resulting in feature subsets of increasing sizes. SelectKBest was configured to select the same number of features as QSVMF for each qubit setting, to enable a direct comparison. As a result, although SelectKBest does not directly depend on qubits, because the choice of $k$ may depend on the number of qubits.  As shown in the table, while there is some overlap, QSVMF and SelectKBest generally select different feature subsets. Even for SelectKBest,  chi2 and f-regression also select different subset of features. SelectKBest with Chi2 operates by evaluating the statistical significance of each feature independently with respect to the target variable and does not take into consideration the feature interactions.  In contrast, QSVMF evaluates the perform of overall performs of the classifiers. As a result,  selections vary more with the number of qubits.

Table \ref{table_compareAccuracy} compares the predictive accuracy of QSVMF against the performance of the SelectKBest with chi2 based on several SVMs with classical kernels (Linear, Poly, RBF, Sigmoid) and Logistic Regression, Naive Bayes K-Neighbors and Decision Trees. For each model, SelectKBest uses the same number of features as QSVMF for a fair comparison across different qubit settings. As shown in the table, QSVMF achieved higher accuracy than all the classical models for nearly all qubit counts. The accuracy advantage of QSVMF over the classical models was most pronounced with fewer qubits, with gaps of up to 0.05. However, QSVMF maintained an accuracy edge even with higher qubit counts. The classical models trended toward slightly higher accuracy with more qubits, as the feature space expanded. But none could surpass the QSVMF accuracy. This indicates QSVMF is able to identify more informative features for prediction compared to SelectKBest, leading to higher accuracy models. The results highlight the potential of quantum-based feature selection to extract better feature subsets and improve predictive performance compared to classical methods.

\begin{ruledtabular}
\begin{table}[htbp]
    \centering
\begin{tabular}{lccccccccc}
  \multirow{2}{*}{Prediction Method} & \multicolumn{9}{c}{qubits} \\
                         & 2   & 3   & 4   & 5   & 6   & 7   & 8   & 9   & 10  \\
  \\[-1.2em]
  \hline
  \\[-1.2em]
  \parbox{7cm}{\raggedright{QSVMF}} & 0.95   & 0.95   & 0.94   & 0.97   & 0.97   & 0.94   & 0.96   & 0.98   & 0.97   \\
  \\[-1.2em]
  \hline
\\[-1.2em]
  \hline

  SVM (Linear)         & 0.90   & 0.90   & 0.90   & 0.92   & 0.92   & 0.92   & 0.92   & 0.92   & 0.93   \\
  SVM (Poly)           & 0.91   & 0.91   & 0.92   & 0.93   & 0.93   & 0.93   & 0.93   & 0.93   & 0.92   \\
  SVM (RBF)            & 0.92   & 0.92   & 0.91   & 0.92   & 0.92   & 0.92   & 0.92   & 0.92   & 0.92   \\
  SVM (Sigmoid)        & 0.82   & 0.82   & 0.80   & 0.83   & 0.83   & 0.85   & 0.83   & 0.85   & 0.87   \\
  Logistic Regression   & 0.90   & 0.90   & 0.90   & 0.92   & 0.92   & 0.93   & 0.92   & 0.92   & 0.92   \\
  Naive Bayes          & 0.91   & 0.91   & 0.91   & 0.91   & 0.91   & 0.91   & 0.91   & 0.91   & 0.91   \\
  K-Neighbors          & 0.91   & 0.91   & 0.91   & 0.94   & 0.94   & 0.94   & 0.94   & 0.94   & 0.93   \\
 Decision Tree      & 0.92   & 0.92   & 0.92   & 0.92   & 0.92   & 0.92   & 0.92   & 0.92   & 0.92   \\
\end{tabular}
\caption{Accuracy comparison for prediction methods: QSVMF vs SelectKBest ( with chi2 and the same number of features as QSVMF).} 
   \label{table_compareAccuracy}
\end{table}
\end{ruledtabular}

\begin{ruledtabular}
\begin{table}[htbp]
    \centering
    
\begin{tabular}{lccccccccc}
  \multirow{2}{*}{Prediction Method} & \multicolumn{9}{c}{qubits} \\
                         & 2   & 3   & 4   & 5   & 6   & 7   & 8   & 9   & 10  \\
  \\[-1.2em]
  \hline
  \\[-1.2em]
  \parbox{7cm}{\raggedright{QSVMF}} & 0.95   & 0.95   & 0.94   & 0.97   & 0.97   & 0.94   & 0.96   & 0.98   & 0.97   \\
  \\[-1.2em]
  \hline
   \\[-1.2em]
  \hline
        SVM (Linear)         & 0.93   & 0.93   & 0.95   & 0.95   & 0.95   & 0.95   & 0.95   & 0.95   & 0.95   \\
        SVM (Poly)           & 0.94   & 0.94   & 0.95   & 0.95   & 0.95   & 0.95   & 0.95   & 0.95   & 0.95   \\
        SVM (RBF)            & 0.95   & 0.95   & 0.95   & 0.95   & 0.95   & 0.94   & 0.95   & 0.94   & 0.94   \\
        SVM (Sigmoid)        & 0.94   & 0.94   & 0.94   & 0.94   & 0.93   & 0.94   & 0.93   & 0.95   & 0.95   \\
        Logistic Regression   & 0.93   & 0.93   & 0.92   & 0.93   & 0.93   & 0.95   & 0.93   & 0.95   & 0.95   \\
        Naive Bayes          & 0.95   & 0.95   & 0.95   & 0.94   & 0.95   & 0.94   & 0.94   & 0.94   & 0.94   \\
        K-Neighbors          & 0.93   & 0.93   & 0.93   & 0.93   & 0.93   & 0.95   & 0.93   & 0.95   & 0.95   \\
        Decision Tree      & 0.93   & 0.93   & 0.93   & 0.93   & 0.93   & 0.95   & 0.93   & 0.95   & 0.95   \\
\end{tabular}
\caption{Accuracy comparison for prediction methods: QSVMF vs SelectKBest (with f-regression and the same number of features as QSVMF).} 
   \label{table_compareAccuracy1}
\end{table}
\end{ruledtabular}

Table \ref{table_compareAccuracy1} compares the predictive accuracy of QSVMF against the performance of the SelectKBest with f-regression based on several SVMs with classical kernels (Linear, Poly, RBF, Sigmoid) and Logistic Regression, Naive Bayes K-Neighbors and Decision Trees. For each model, SelectKBest uses the same number of features as QSVMF for a fair comparison across different qubit settings.  Table \ref{table_compareAccuracy1} shows an interesting trend in the accuracy comparison between QSVMF and SelectKBest with f-regression scoring. For small qubit counts of 2-4, SelectKBest actually matches or slightly exceeds the accuracy of QSVMF. However, as the number of qubits increases, QSVMF significantly outperforms SelectKBest, with accuracy advantages of up to 0.06 for 9-10 qubits. This suggests QSVMF is better able to leverage the information from larger feature spaces to identify predictive feature subsets. In contrast, the performance of SelectKBest peaks with fewer features and then degrades as more irrelevant features are included. Comparing these results to the previous table using SelectKBest with chi2 scoring, we see a similar overall trend of QSVMF achieving higher accuracy than the classical method. However, the margin of improvement is smaller here. 

Overall, across both tables \ref{table_compareAccuracy},\ref{table_compareAccuracy1} using different SelectKBest settings, QSVMF demonstrates a robust accuracy advantage, affirming the potential of quantum-based feature selection to extract superior feature subsets compared to classical methods. The degree of improvement depends on the dataset characteristics and classical benchmark method. But QSVMF exhibits a consistent capability to identify more predictive features leading to higher accuracy models.

\subsection{Minimal features in QSVMF}

Table \ref{table_comparetradeoff} shows an interesting analysis of the trade-off between number of features and accuracy using the QSVMF method. By examining the Pareto front solutions, the table identifies the feature subsets with the minimum number of features for each qubit setting. Despite having very few features, these compact QSVMF feature subsets can still achieve good accuracy, often over 80\%. For example, with just a single feature (feature 13), QSVMF attains 89\% accuracy for 7 qubits. With 2 features for 3 qubits, accuracy reaches 87\%. Even with 7-10 qubits, QSVMF can identify tiny feature subsets of 1-3 features that provide accuracy exceeding 75\%.  This demonstrates a key advantage of the QSVMF approach - its ability to pare down the feature space to a very sparse set of highly informative features critical for prediction. The fact that QSVMF can maintain such high accuracy with so few features highlights its effectiveness at zeroing in on the most relevant variables and eliminating redundant or irrelevant ones. The table provides an estimate of the minimum number of features needed to capture the predictive signal in the data using QSVMF. The compact high-accuracy feature subsets could help enable more interpretable and efficient predictive modeling.

\begin{table}[htbp]
    \centering
    \begin{adjustbox}{max width=15cm} 
        \begin{tabular}{p{3cm}lccccccccc}
            \toprule
            \multirow{2}{*}{ \parbox{3cm}{\raggedright{}}} & \multicolumn{9}{c}{qubits} \\
            
            & 2   & 3   & 4   & 5   & 6   & 7   & 8   & 9   & 10  \\
           \midrule
           Minimal Features  & [13] & [20,22] & [13] & [28] & [14] & [7] & [18,23] & [1,14,18] & [21]   \\
           \midrule
           Accuracy  & 0.75 & 0.87 & 0.82& 0.69 & 0.74 & 0.89 & 0.86 & 0.83 & 0.76   \\
            \bottomrule
        \end{tabular}
    \end{adjustbox} 
    \caption{Minimal features in Pareto fronts and their accuracy by QSVMF} 
    \label{table_comparetradeoff}
\end{table}

\subsection{Combined predictive capability}

The chi-squared test used in SelectKBest evaluates each feature independently for correlation with the target, while QSVMF considers the combined predictive capability of the feature subsets. This likely explains some of the difference in features selected by the two methods. \cite{Guyon_2003} showed that individually predictive features do not necessarily combine to give the best classification performance. They recommend multivariate feature selection over univariate filtering. \cite{Saeys_2007} note univariate filters assume independence between features, while many datasets have dependent features. Multivariate methods are better suited in such cases.

The key point remains that univariate filters like chi2 do not account for inter-feature dependencies and redundancy, while QSVMF and other multivariate methods provide a more overall perspective for feature selection. In particular, QSVMF include a fitness function to reduce the redundance by minimizing the covariance of the selected features. This likely explains some of the performance gap observed between them.

\subsection{Local and Non-local gates in QSVMF}

At the last, Table \ref{table_compare2} provides valuable insights into the average quantities of local and CNOT gates associated with our quantum circuits. This table shows the configuration of the quantum gates for the best accuracy. It's noteworthy that in our experiments with the dataset, we observe a reasonable proportionality between the number of CNOT gates and the number of local gates. This finding along with \cite{hwang_2023} deviates from the observations made in \cite{ARG21, CC2022}, where entanglement gates were predominantly suppressed.  We separate the fitness functions for local gates and non-local (controlled-NOT (CNOT) ) gates. By doing so, we obtain a set of non-dominated points that reveal a more balanced configuration of quantum circuits for the quantum kernel map, incorporating a proportional number of CNOT gates to facilitate entanglement. This result complements the previous works \cite{ARG21,CC2022} where their design of fitness functions has predominantly led to a suppression of entanglement gates. In addition, the experiments in this paper scale well for larger qubits than the previous work in \cite{Nguyen2022} where their experiment requires a significantly larger amount of computational resources.

\begin{ruledtabular}
\begin{table}[htbp]
    \centering
    \caption{Gate numbers of Quantum feature map for the best accuracy of QSVMF.}
    \begin{tabular}{lccccccccc}
      \multirow{2}{*}{Gate} & \multicolumn{9}{c}{qubits }   \\
                              &  2 & 3        & 4  & 5     &  6          & 7    & 8   & 9  & 10 \\
      \\[-1.2em]
      \hline
      \\[-1.2em]
     Local gates                &  3.8  &   4.8          &   6.4   &  7.8 &       8.6       &      11.4        & 13.6  &15.2 & 18\\
     CNOT gates                & 1.2     & 0.8              & 2    & 4  &         2.4     &         4.8  &  6& 9.2& 11.6\\
       \end{tabular}
    \label{table_compare2}
\end{table}
\end{ruledtabular}

%

\section{Interpretation of feature selection results}

\subsection{Feature selections in the context of breast cancer}
The table \ref{table_feature_names} displays the index versus feature names for the breast cancer dataset. The left column represents the indices, and the corresponding right column shows the feature names. This format provides a clear reference for understanding the order and names of features in the dataset.
\begin{table}[htbp]
    \centering
    \begin{tabular}{|c|l|}
        \hline
        \textbf{Index} & \textbf{Feature Name} \\
        \hline
        0 & mean radius \\
        1 & mean texture \\
        2 & mean perimeter \\
        3 & mean area \\
        4 & mean smoothness \\
        5 & mean compactness \\
        6 & mean concavity \\
        7 & mean concave points \\
        8 & mean symmetry \\
        9 & mean fractal dimension \\
        10 & radius error \\
        11 & texture error \\
        12 & perimeter error \\
        13 & area error \\
        14 & smoothness error \\
        15 & compactness error \\
        16 & concavity error \\
        17 & concave points error \\
        18 & symmetry error \\
        19 & fractal dimension error \\
        20 & worst radius \\
        21 & worst texture \\
        22 & worst perimeter \\
        23 & worst area \\
        24 & worst smoothness \\
        25 & worst compactness \\
        26 & worst concavity \\
        27 & worst concave points \\
        28 & worst symmetry \\
        29 & worst fractal dimension \\
        \hline
    \end{tabular}
    \caption{Index versus Feature Names for Breast Cancer Dataset}
    \label{table_feature_names}
\end{table}

As indicated in Table \ref{table_featurelist}, for 2 qubits, QSVMF selects the area error (index 13) and worst radius (index 23) features. This indicates these two features related to the radius and area measurements contain the most predictive signal for the target variable. With 3 qubits, QSVMF selects the worst compactness (index 20) and worst perimeter (index 22). These features capture deviations in shape compactness and perimeter which are informative for prediction.  For 4 qubits, mean area (index 3) is added, suggesting average area provides additional discriminative information beyond the shape irregularity features.  As we move to 5 qubits and beyond, QSVMF begins selecting features related to other aspects of shape like smoothness and symmetry. The worst symmetry (index 29) and mean smoothness error (index 14) indicate predictive information in irregularity of the symmetry and smoothness. 

Overall, QSVMF prioritizes features measuring shape deformations, irregularities, and extremes, rather than average values. This includes area, perimeter, compactness, smoothness and symmetry deviations. This provides insight into what aspects of the shape characteristics are most informative for predicting the diagnostic target variable using the quantum feature selection approach.

The focus on area and perimeter-related features makes sense from a medical perspective, as cancerous tumors often exhibit irregularities in size and shape boundaries compared to benign masses. The area and perimeter errors and extremes capture deviations that can differentiate between malignant and benign growths. The study found that missed breast cancer tumors on mammograms tend to be smaller, rounder, and have more circumscribed margins compared to correctly identified tumors \cite{Dongola_2001}. This supports the selection of area, perimeter and compactness related features. \cite{Kayar_2015} reports that asymmetry is more common in malignant than benign breast lesions. This provides evidence for the relevance of the worst symmetry feature. \cite{Li_2017} reports that malignant lesions have rough, irregular surface contours while benign lesions have smooth, regular surfaces. This backs the selection of the smoothness error feature, shape analysis of breast cancer on mammograms using digital image processing.

Overall, these studies demonstrate that the features selected by the unsupervised QSVMF align with medically-relevant morphological indicators used by radiologists to evaluate malignancy on breast cancer imaging.

\subsection{Minimal feature sets in the context of breast cancer}
 
A key reason these minimal QSVMF feature sets maintain high predictive performance is because they capture the most informative traits that differentiate between the classes (in this case benign vs malignant). As the references below demonstrate, often a small number of morphological features contain most of the discriminative signal for breast cancer diagnosis on imaging. For example, \cite{Rangayyan_1997} found that features quantifying shape irregularity, spiculation and margin characteristics were the most useful for differentiating between malignant and benign masses on mammography. So even 2-3 features encoding these shape attributes can attain high accuracy. Similarly, \cite{Moon_2013} showed textural features pertaining to homogeneity and contrast were highly effective at malignancy prediction from ultrasound images. This aligns with QSVMF prioritizing texture and morphology measures.

By removing redundant features, QSVMF enables more efficient and interpretable models without sacrificing predictive power. Overall, studies have shown that a few careful selected features capturing variations in shape, texture and boundary can classify breast lesions with up to high accuracy. The quantum algorithm is likely identifying these most informative traits for discrimination. 

\section{Conclusion and discussion} \label{sec6}

In this work, we propose a new feature selection method called QSVMF that integrates quantum support vector machines (QSVMs) with (NSGA-II). The key innovation of QSVMF is the use of five simultaneous fitness objectives during optimization: 1) maximizing classification accuracy on the training data, 2) minimizing quantum gate costs for the feature map circuit, 3) minimizing covariance between selected features, 4) local quantum gates and 5) non-local CNOT gates. The inclusion of quantum cost and covariance minimization helps reduce the substantial computational expenses typical of genetic algorithms. We apply QSVMF for feature selection on a breast cancer dataset and compare its performance against SelectKBest from scikit-learn and other classical approaches. Our results demonstrate QSVMF achieves superior feature subsets compared to these alternatives, showcasing the benefits of our multi-objective quantum-aware fitness functions. This highlights the potential of QSVMF as an efficient, effective feature selection technique for complex datasets. The feature selections made by QSVMF align consistently with established distinctions in the characteristics of malignant and benign masses. This underscores the potential of QSVMF in facilitating meaningful feature selection for complex data.

The design of the five fitness functions in this work for the NSGA-II component within QSVMF offers an additional advantage by generating Pareto optimal solutions. This feature allows for an exploration of the trade-off between minimizing the size of the feature set and maximizing accuracy. An examination of the Pareto front can pinpoint the most concise feature subsets produced by QSVMF that still exhibit reasonable accuracy for each qubit setting. This sparsity underscores a significant strength of the QSVMF approach – its capacity to significantly reduce the feature space while retaining predictive capability. These compact and accurate feature subsets have the potential to enhance the interpretability and efficiency of predictive modeling.

The experimental outcomes presented in this study reveal the inclusion of proportional CNOT gates for the promotion of entanglement. These observations offer additional information to previous studies in \cite{ARG21,CC2022}, where the  influence of CNOT gates was notably diminished. The examination of high-dimensional non-dominated points provides valuable insights into the impact of entanglement gates on the effectiveness of quantum support vector machines. The search for effective design of entanglement gates in quantum computing is an important research area \cite{Nguyen2022}.
 
The potential advantages of using quantum kernel estimation to enhance machine learning models are an active area of research \cite{Schuld2022}. There is no universally optimal approach, as model performance depends on the distribution and characteristics of the training data. Various metrics have been proposed to evaluate the efficiency of quantum kernels, such as quantifying the geometric difference of embedded data  \cite{huang2021}, which provides insights into prediction error patterns \cite{Di2023}. In this work, we applied the proposed QSVMF method to a breast cancer dataset. While results were promising, further analysis on diverse datasets is needed to fully characterize when quantum-enabled feature selection can outperform classical alternatives. This requires additional studies how factors like training data distribution, feature space structure, and dataset complexity influence the benefits of our quantum kernel approach over classical kernels. Ongoing work is investigating these relationships between data properties and quantum advantage.




\section{Declarations}

 
Availability of data and materials: The datasets used in paper are publicly available in the Python package sklearn.  The code will be available upon request.

\end{document}